\documentclass[pra,showpacs,twocolumn]{revtex4}
\usepackage{graphicx}
\usepackage{epstopdf}

\begin{document}

\title{Effective Hamiltonian Theory and Its Applications in Quantum Information*}
\author{Daniel F. V. James}
\affiliation{Department of Physics, University of Toronto, 60, St. George St., Toronto, Ontario, Canada M5S 1A7. \email{dfvj@physics.utoronto.ca}}
\author{Jonathan Jerke}
\affiliation{Yale University, Physics Department, P.O. Box 208120, New Haven, CT 06520, USA.\email{jonathan.jerke@yale.edu}}
\begin{abstract}
This paper presents a useful compact formula for deriving an effective Hamiltonian describing the time-averaged dynamics of detuned quantum systems. The formalism also works for ensemble-averaged dynamics of stochastic systems. To illustrate the technique we give examples involving Raman processes, Bloch-Siegert shifts and Quantum Logic Gates.\\
\begin{center}* To be published in {\em Canadian Journal of Physics}.\end{center}
\end{abstract}

\maketitle

\section{Introduction}
\setcounter{equation}{0}

In analyzing the dynamics of quantum mechanical systems one is often confronted by Hamiltonian terms which are rapidly varying in time.  And yet such off-resonant terms, which can require elaborate theoretical or numerical handling,  often end up having rather mundane effects.  For example, consider a weak, far off resonant harmonic perturbation is applied to a two-level system (Fig.1).  Its interaction picture Hamiltonian is 
\begin{equation}
\hat{H}=\frac{\hbar \Omega}{2}\left\{
|2\rangle\langle 1|\exp(-i\Delta t)+
|1\rangle\langle 2|\exp(i\Delta t)\right\},
\label{alpha}
\end{equation}
where $\Omega$ is the Rabi frequency, which characterizes the strength of the interaction (and is assumed to be small compared with the detuning $\Delta$). When we calculate the dynamics of such a system in the Bloch-vector representation \cite{alleneberly}, we find a very rapid precession of the Bloch vector about the effective torque, which very nearly points along the z-axis. And yet, the {\em effect} of such a perturbation is a slight change of the resonance frequency of the two levels (know as the ``A.C. Stark shift" in quantum optics), described by an effective Hamiltonian of the form
\begin{equation}
\hat{H}_{eff}=-\frac{\hbar \Omega^2}{4 \Delta}\left\{|2\rangle\langle 2|-|1\rangle\langle 1|\right\}.
\label{beta}
\end{equation}
This begs the question: is there a straightforward way to transform the rapidly oscillating Hamiltonian eq.(\ref{alpha}) into a time averaged effective Hamiltonian eq.(\ref{beta}) which accounts for the actual effect of the interaction?

\begin{figure}[!b]
\vspace{-3mm}
\includegraphics[width=35mm]{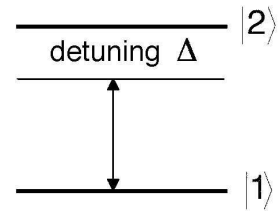}
\vspace{-2mm}
\caption{A two level system with a detuned interaction.}
\label{geofig}
\end{figure}

\section{Derivation of the Effective Hamiltonian}
\setcounter{equation}{0}

In the interaction picture, the state of a quantum system $|\psi\left(t\right)\rangle$ evolves according to the following formula, 
\begin{equation}
|\psi\left(t\right)\rangle = \hat{U}\left(t,t_0\right)|\psi\left(t_0\right)\rangle
\label{HounslowWest}
\end{equation}
where $ \hat{U}\left(t,t_0\right)$ is the unitary time evolution operator which obeys the equation
\begin{equation}
i \hbar \frac{\partial}{\partial t} \hat{U}\left(t,t_0\right) =\hat{H}_I \left(t \right) \hat{U}\left(t,t_0\right),
\label{HounslowCentral}
\end{equation}
where $\hat{H}_I \left(t \right)$ is the interaction Hamiltonian.  We have assumed the interaction picture here and throughout this paper. 
\footnote{In general, the definition of the Interaction picture is not unique, because the splitting of  a Hamiltonian into an unperturbed part and a perturbed part is rather arbitrary; one can avoid this problem by defining the unperturbed part of the Hamiltonian as $\hat{H}_0=\lim_{T\rightarrow\infty}\frac{1}{2 T}\int^T_{-T}\hat{H}\left(t'\right) dt'$.}

We are considering the coarse grained or time averaged dynamics of the system, defined formally by the following averaging procedure for some operator $\hat{\mathcal{O}}\left(t\right)$ :
\begin{equation}
\overline{\hat{\mathcal{O}}\left(t\right)}=\int_{-\infty}^{\infty}f\left(t-t'\right)\hat{\mathcal{O}}\left(t'\right)dt',
\label{Osterley}
\end{equation}
where the function $f\left(t\right)$ is real valued, has unit area, and is assumed to act as a low-pass filter (so that high frequency terms disappear from the average).  Time averages defined in this way are well-known in describing the dynamics of quantum systems: the {\em Rotating Wave Approximation}  \cite{alleneberly} is routinely used to discard various high frequency components in Hamiltonians of systems.  Our time averaging is nothing more than a generalization of this approximation.
Causality requires $f\left(t\right)=0$ if $t<0$, although this is not a strict requirement of the theory.  As is well known, the use of the rotating wave approximation leads to some non-causal predictions \cite{Milonni}; this will also be true of our generalization of the the R.W.A.: one cannot use these techniques in calculations where causal behaviour is critical. Thus, while it may be strictly correct to use a causal averaging kernel for coarse-gained or time averaged dynamics, a non-causal function (e.g. a Gaussian function peaked at $t=0$, and therefore non-zero for $t<0$) may be easier to manipulate mathematically. In what follows we will avoid where possible making a specific choice of $f\left(t\right)$, instead using generic properties of time averages.  Using integration by parts, it can then be shown that the time derivative of the average is the average of the time derivative, i.e. $\partial \overline{\hat{\mathcal{O}}\left(t\right)}/\partial t = \overline{\partial \hat{\mathcal{O}}\left(t\right)/\partial t }$.

The effective Hamiltonian may now be defined straightforwardly by considering the time derivative of 
the {\em time-averaged} time-evolution operator, viz.,
\begin{equation}
i \hbar \frac{\partial}{\partial t} \overline{\hat{U}\left(t,t_0\right)} =\hat{\cal H}_{eff} \left(t \right) \overline{ \hat{U}\left(t,t_0\right)}
\label{BostonManor}
\end{equation}

If we apply the averaging procedure discussed in eq.(\ref{Osterley}) to eq.(\ref{beta}) and compare it with the eq.(\ref{BostonManor}) we obtain
\begin{equation}
\hat{\cal H}_{eff} \left(t \right) \overline{\hat{U}\left(t,t_0\right)}=
\overline{\hat{H}_I \left(t \right) \hat{U}\left(t,t_0\right)},
\label{Northfields}
\end{equation}
which immediately leads to the formula
\begin{equation}
\hat{\cal H}_{eff} \left(t \right) =
\left\{\overline{\hat{H}_I \left(t \right) \hat{U}\left(t,t_0\right)} \right\}
\,
\left\{\overline{\hat{U}\left(t,t_0\right)} \right\}^{-1}.
\label{gamma}
\end{equation}
Note that, although $\hat{U}\left(t,t_0\right)$ is unitary, the time averaged operator is in general not.  Thus this expression Eq.({\ref{gamma}) has the serious drawback that it is not Hermitian.  The origin of this non-unitary evolution is easy to see: we have, in effect, divided our Hilbert space into a low-frequency part, and a high-frequency part; time averaging is equivalent to performing a partial trace over the high-frequency part.  Performing a partial trace over one part of the Hilbert space is a classic paradigm for describing non-unitary dynamics.  One can carry out a procedure entirely analogous to the derivation of the master equation (see for example \cite{GSA}) for this particular case, leading to the result that the effective Hamiltonian for the unitary part of the evolution is uniquely given by the Hermitian part of $\hat{\cal H}_{eff} \left(t \right)$, i.e.
\begin{equation}
\hat{ H}_{eff} \left(t \right) = \frac{1}{2}\left\{ \hat{\cal H}_{eff} \left(t \right)+\hat{\cal H}_{eff} \left(t \right)^{\dagger}\right\}.
\end{equation}

We can obtain an expression for $\hat{\cal H}_{eff} \left(t \right)$ using the standard expansion for the time-ordered evolution operator, 
$\hat{U}\left(t,t_0\right)=\hat{T}\exp\left[-i\int^{t}_{t_0}\hat{H_I}\left(t'\right)dt'/\hbar\right] $.  Keeping terms to the second order of the Hamiltonian $\hat{H_I}\left(t\right)$ we obtain
\begin{equation}
\hat{\cal H}_{eff} \left(t \right) = \overline{\hat{H_I}\left(t\right)}+\overline{\hat{H_I}\left(t\right)\hat{U}_1\left(t\right)}-
\overline{\hat{H_I}\left(t\right)}\,\,\overline{\hat{U}_1\left(t\right)}+\ldots,
\end{equation}
where 
\begin{equation}
\hat{U}_1\left(t\right) =\frac{1}{i\hbar} \int^{t}_{t_0}\hat{H_I}\left(t'\right)dt'.
\end{equation}
Since the time-averaging function $f(t)$ was assumed to be real, it follows that the Hermitian adjoint and time averaging commute; using the fact that $\hat{H_I}\left(t\right)^{\dagger}=\hat{H_I}\left(t\right)$ and $\hat{U}_1\left(t\right)^{\dagger}=-\hat{U}_1\left(t\right)$, we find that
\begin{equation}
\hat{ H}_{eff} \left(t \right) = \overline{\hat{H}\left(t\right)}+
\frac{1}{2}\left(
	\overline{\left[\hat{H}\left(t\right),\hat{U}_1\left(t\right)\right]}
	-\left[\overline{\hat{H}\left(t\right)},\overline{\hat{U}_1\left(t\right)}\right]
\right).
\label{delta}
\end{equation}

\section{Harmonic Time dependence}
\setcounter{equation}{0}

It is often the case that one deals with systems undergoing a perturbation with an harmonic time dependence, at one or more frequency. Examples of this include an atomic or molecular system interacting with one or more laser beams; or an electron or nuclear spin interacting with an oscillating magnetic field; it is not difficult to think of other cases.  In such systems the interaction Hamiltonian will have the following form:
\begin{equation}
\hat{H_I}\left(t\right) = \sum_{n=1}^{N}
\hat{h}_n \exp{\left(-i \omega_{n}t\right)}+\hat{h}_n^{\dagger} \exp{\left(i \omega_{n}t\right)},
\label{foxtrot}
\end{equation}
Where $N$ is the total number different harmonic terms making up the interaction Hamiltonian.  We assume (without loss of generality) that $\omega_{n}>0$, and that $\omega_{1}\le\omega_{2}\le\ldots\le\omega_{N}$. We will assume that the range of frequencies present in the interaction is small, i.e. that $\max_{m,n}|\omega_{n}-\omega_{n}|\equiv \omega_{N}-\omega_{1}<<\omega_{1}$.

In this case, we can immediately calculate the operator $\hat{U}_1\left(t\right)$:
\begin{equation}
\hat{U}_1\left(t\right) =\hat{V}_1\left(t\right)-\hat{V}_1\left(t_0\right).
\end{equation}
where
\begin{equation}
\hat{V}_1\left(t\right) =\sum_{n=1}^{N} \frac{1}{\hbar \omega_n} \left(\hat{h}_n e^{-i \omega_{n}t}-\hat{h}_n^{\dagger} e^{i \omega_{n}t}\right).
\end{equation}
If the initial time $t_0$ is in the distant past 
\footnote{Specifically, we require that $(t-t_0)>>\tau$, where $\tau$ is the characteristic width of the time-averaging kernel $f(t)$.  This requirement means that the theory cannot be expected to be applicable for very short time periods.}, 
we may assume that $\overline{\hat{V}_1\left(t_0\right)}=\hat{V}_1\left(t_0\right)$, and immediately the term cancels from 
eq.(\ref{delta}), implying that
\begin{equation}
\hat{ H}_{eff} \left(t \right) = \overline{\hat{H_I}\left(t\right)}+
\frac{1}{2}\left(
	\overline{\left[\hat{H_I}\left(t\right),\hat{V}_1\left(t\right)\right]}
	-\left[\overline{\hat{H_I}\left(t\right)},\overline{\hat{V}_1\left(t\right)}\right]
\right).
\label{epsilon}
\end{equation}
We will assume that the time average procedure acts as a low-pass filter.  In other words, we are only going to concern ourselves with dynamic processes occurring at low frequencies; all processes oscillating at frequencies $\omega_{1}$ and above are assumed to average out to zero.  Mathematically, we assume that, for all values of $m,n$
\begin{eqnarray}
\overline{\exp\left({\pm i \omega_{n}t}\right)}&=&0 \label{mabel}\\
\overline{\exp\left({\pm i [\omega_{n}+\omega_{m}]t}\right)}&=&0 \label{josephine}\\
\overline{\exp\left({\pm i [\omega_{n}-\omega_{m}]t}\right)}&=&\exp\left({\pm i [\omega_{n}-\omega_{m}]t}\right).
\label{pinafore}
\end{eqnarray}
Using these assumptions, one immediately finds that 
\begin{eqnarray}
\overline{\hat{H}_I\left(t\right)}&=&0\\
\overline{\hat{V}_1\left(t\right)}&=&0.
\label{penzance}
\end{eqnarray}
Substituting from eqs.(\ref{pinafore}) and (\ref{penzance}) into eq.(\ref{epsilon}) one therefore finds, after a little bit of algebra, the following compact formula for the effective Hamiltonian:
\begin{equation}
\hat{ H}_{eff} \left(t \right) =
\sum_{m,n=1}^{N}
 \frac{1}{\overline{\hbar \omega}_{mn}}
  \left[\hat{h}_m^{\dagger},\hat{h}_n\right] \exp\left({ i [\omega_{m}-\omega_{n}]t}\right),
\label{zeta}
\end{equation}
where $\overline{\omega}_{mn}$ is the harmonic average of $\omega_{m}$ and $\omega_{n}$, viz.,
\begin{equation}
\frac{1}{\overline{\omega}_{mn}}=\frac{1}{2}\left(\frac{1}{\omega_{m}}+\frac{1}{\omega_{n}}\right).
\end{equation}
This formula eq.(\ref{zeta}) is the principal result of this paper.  It was derived in a different manner by one of us a few years ago \cite{dfvj}, in a paper dealing principally with interactions between laser and trapped ions, and the applicability of such systems as a possible implementation of a quantum computer.  Since then various authors have employed the technique \cite{Plenio,Molmer,Mathis}, prompting us to undertake a more thorough investigation of the formula's applicability. To summarize, the assumptions that have led from eq.(\ref{foxtrot}) to eq.(\ref{zeta}) are (i) the interaction is sufficiently weak that we can we can terminate the series expansion of the time evolution operator after two terms; (ii) that the system is not responsive to high frequency stimulation, allowing us to make the time-averaging assumptions embodied in eq.(\ref{mabel})-eq.(\ref{pinafore}); and (iii) the interaction has been occurring for a sufficiently long time that transient effects can be neglected.  Note however that the theory is {\em not} a perturbation theory; if the effective Hamiltonian is sufficiently simple that Schr\"odinger's equation can be solved , such a solution would embody all powers of the original interaction Hamiltonian.  The solution of the propagator in these circumstances would be equivalent to a partial resumption of a Dyson series for the propagtor.

Examinations of time-averaged dynamics have been presented by other authors. In particular, Avan {\em et al.} \cite{HeffCT} examined the problem using the Schr\"odinger picture, and obtained an expression involving sums over matrix elements which is related to ours.  More recently, the notion of an effective Hamiltonian for specific models of non-linear optical interactions has been presented by Klimov et al. \cite{Klimov}. Furthermore, there is a considerable body of work in the nuclear magnetic resonance literature dealing with the notion of {\em averaged Hamiltonians}, which is a related concept (see for example \cite{AHT}).  However, to our knowledge, the compact formula eq.(\ref{zeta}) given here was not employed by any of these authors. We will now give some examples of its utility.

\section{Examples}
\setcounter{equation}{0}

\subsection{The A.C. Stark Shift}
As discussed above, the A.C. Stark shift is an effective frequency shift experienced by a two level system due to the application of a far off resonance harmonic perturbation.  As mentioned above the Hamiltonian describing this system is 
\begin{equation}
\hat{H}_I\left(t\right)=\frac{\hbar \Omega}{2}\left\{|2\rangle\langle 1|\exp(-i\Delta t)+|1\rangle\langle 2|\exp(i\Delta t)\right\}.
\end{equation}
Immediately we can make the identification $\hat{h}_1=\left( \hbar \Omega/2\right)|2\rangle\langle 1|$ and $\omega_1=\Delta$.  Applying eq.(\ref{zeta}), and evaluating the commutator, we quickly find that 
\begin{equation}
\hat{H}_{eff}=-\frac{\hbar \Omega^2}{4 \Delta}\left\{|2\rangle\langle 2|-|1\rangle\langle 1|\right\}.
\end{equation}
Note that, if the detuning $\Delta$ were a negative quantity, we would have had to identify $\hat{h}_1=\left( \hbar \Omega/2\right)|1\rangle\langle 2|$, since we have assumed  $\omega_1$ is positive; carrying through the calculation results in a change of sign of the effective Hamiltonian, as one would expect.

A related effect is the Bloch-Siegert shift \cite{alleneberly}, which is a small shift in the resonance of a two level system which occurs due to the very high frequency components of the dipolar interaction, normally neglected under the Rotating Wave Approximation.  If one introduced the R.W.A. in a systematic way by employing some time averaging procedure, one could carry out an analysis analogous to that presented here, resulting in a small frequency shift term. One could also argue that the Lamb shift could be ascribed to an A.C. Stark shift caused by the vacuum fluctuations of the electromagnetic field, although in this case one would have to deal with the on-resonant components in a different manner.

\subsection{Raman transitions}
As the next level of complexity, consider the case of a three level system interacting with two time-harmonic terms (e.g. fig.2)

\begin{figure}[!b]
\vspace{-3mm}
\includegraphics[width=55mm]{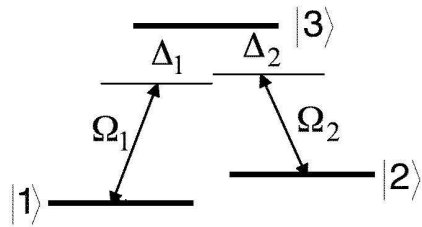}
\vspace{-2mm}
\caption{A three level system with two detuned interactions.}
\label{ramanfig}
\end{figure}

In this case the interaction Hamiltonian is 
\begin{eqnarray}
\hat{H}_I\left(t\right) &=& \frac{\hbar \Omega_1}{2}|3\rangle\langle 1|\exp(-i\Delta_1 t)\nonumber\\
&&+\frac{\hbar \Omega_2}{2}|3\rangle\langle 2|\exp(-i\Delta_2 t) + h.a.
\end{eqnarray}
where $h.a.$ stands for the Hermitian adjoint of the preceding terms.  Again, it is straightforward to make the identification $\hat{h}_1= (\hbar \Omega_1/2)|3\rangle\langle 1|$, $\hat{h}_2= (\hbar \Omega_2/2)|3\rangle\langle 2|$, $\omega_1=\Delta_1$ and $\omega_2=\Delta_2$; substituting into eq.(\ref{zeta}) and evaluating the commutators (four, in total), we find the effective Hamiltonian given by
\begin{eqnarray}
\hat{H}_{eff}&=&-\frac{\hbar \Omega_1^2}{4 \Delta_1}\left\{|3\rangle\langle 3|-|1\rangle\langle 1|\right\}-
\frac{\hbar \Omega_2^2}{4 \Delta_2}\left\{|3\rangle\langle 3|-|2\rangle\langle 2|\right\} \nonumber \\
&&+\frac{\hbar \Omega_1\Omega_2}{4 \overline{\Delta}}
\left(|1\rangle\langle 2| \exp\left\{i(\Delta_1-\Delta_2) t\right\}\right.\nonumber\\
&&\left.-|2\rangle\langle 1| \exp\left\{-i(\Delta_1-\Delta_2) t\right\}\right).
\end{eqnarray}
The first two terms can be identified with A.C. Starks shifts associated with the two lasers; the second pair of terms represent transitions of population between the levels $|1\rangle$ and $|2\rangle$.  Despite there being no direct interaction between these two levels (indeed, for dipole interactions, if the transitions $|1\rangle\to|3\rangle$ and $|2\rangle\to|3\rangle$ are allowed, then $|1\rangle\to|2\rangle$ is forbidden), their population can be induced to oscillate coherently.  These are the well-known Raman transitions, which play an important role in many areas of atomic and molecular interactions with light.

\subsection{Quantum A.C. Stark Shifts}
Consider the situation shown schematically in figure 3.  An ion is confined in an harmonic well, and is free to oscillate in the $z$-direction, and is interacting with a laser.  Let us assume that the internal degrees of freedom of the ion can be modeled as a two level system, while the external degrees of freedom (i.e. its displacement from its equilibrium position in the trap) will be described by the standard quantum harmonic oscillator ladder of equally spaced quantum states.  

\begin{figure}[!b]
\vspace{-3mm}
\includegraphics[width=75mm]{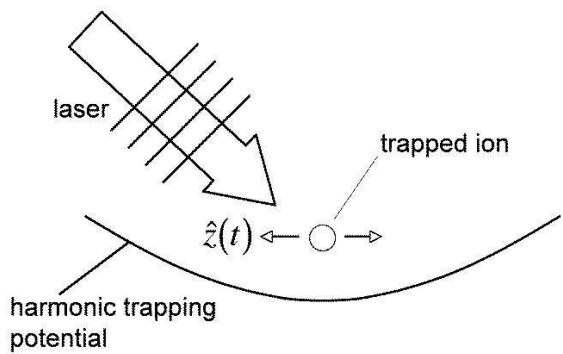}
\vspace{-2mm}
\caption{A trapped ion, free to oscillate in the one direction, and interacting with a laser field.}
\label{ionfig}
\end{figure}

The Hamiltonian describing this interaction is as follows
\begin{equation}
\hat{H}_I\left(t\right) = \frac{\hbar \Omega}{2}|2\rangle\langle 1|{\rm e}^{i k_z \hat{z}(t)-i\Delta t} + h.a..
\end{equation}
In this case $\Delta$ is the detuning between the laser and the internal resonance of the ion, $k_z$ is the component of the wavevector of the laser along the z-direction and the operator $\hat{z}(t)$ is given by
\begin{equation}
k_z \hat{z}(t)=\eta\left(\hat{a} {\rm e}^{-i\omega_0 t}+\hat{a}^\dagger {\rm e}^{i\omega_0 t}\right),
\end{equation}
where $\eta$, Lamb-Dicke parameter, is a dimensionless coupling constant, $\hat{a}^\dagger$ and $\hat{a}$ are the creation and annihilation operators for the harmonic oscillator, and $\omega_0$ is the trap frequency (see \cite{dfvj2} for a more  detailed discussion).  Usually the dimensionless parameter $\eta$ is small, hence we can make the ``Lamb-Dicke" approximation, $\exp{i k_z \hat{z}(t)}\approx 1+i k_z \hat{z}(t)$. This allows us to make the identification of three harmonic terms in this Hamiltonian
\begin{eqnarray}
\hat{h}_1&=&\frac{i\hbar \eta \Omega}{2}|2\rangle\langle 1|\hat{a}^\dagger\\
\hat{h}_2&=&\frac{\hbar \Omega}{2}|2\rangle\langle 1|\\
\hat{h}_3&=&\frac{i\hbar \eta \Omega}{2}|2\rangle\langle 1|\hat{a},
\end{eqnarray}
with $\omega_1=\Delta-\omega_0$,  $\omega_2=\Delta$ and  $\omega_3=\Delta+\omega_0$. In this case the effective Hamiltonian is
\begin{eqnarray}
\hat{H}_{eff}&=&-\frac{\hbar \Omega^2}{4 \Delta}
\left(
1+2\frac{\eta^2 \Delta^2}{\Delta^2-\omega_0^2}\left(\hat{n}+1/2\right),
\right)\nonumber\\
&&\times
\left\{|2\rangle\langle 2|-|1\rangle\langle 1|\right\}
\end{eqnarray}
where $\hat{n}=\hat{a}^\dagger\hat{a}$ is the number operator for the harmonic well.
Once again, this effective Hamiltonian has the form of a frequency shift, but in this case the shift is dependent on the population of the the harmonic oscillator.  Such `quantum' A.C. Stark have been studied before \cite{DM}; indeed, since one may, using an appropriately detuned laser pulse, coherently map the internal degrees of freedom of an ion onto the external degrees of freedom, interactions of this kind have been proposed for use as quantum gates \cite{SJM, FSK}.  

\subsection{Quantum gates}
Finally we consider the application of our formula to the derivation of quantum gate operations (this discussion first appeared in our earlier paper \cite{dfvj} in which eq.(\ref{zeta}) was first derived).
Quantum gates between two-level qubits based on trapped ions and employing off-resonant interactions were proposed by M\o lmer and S\o rensen a few years ago
 \cite{sorensen1,sorensen2}.  These proposals have the distinct advantage that they perform well even with ions in any mixed state of the ions' collective oscillation modes, and were subsequently used to perform the deterministic entanglement of four ions \cite{NISTfour}. 
A laser field with {\em two} spectral components, detuned equally to the red and to the blue of the 
ions' atomic resonance frequency is applied to a two ions in
the trap.  In the Lamb-Dicke approximation, the interaction is described by the following
Hamiltonian:
\begin{eqnarray}
\hat{H}_{I} \left(t\right)&=&\frac{\hbar\Omega}{2}\hat{J}^{(+)}
\left\{1+i\eta
\left(
\hat{a}e^{-i\omega_{0}t}+\hat{a}^{\dagger}e^{i\omega_{0}t}\right)
\right\}
\cos(\Delta t)\nonumber\\
&&\,\,\,\,\,\,\,\,\,\,\,\,\,\,\,\,\,\,\,\,\,\,\,\,\,\,\,\,\,\,\,\,\,\,\,\,\,\,\,\,\,\,\,\,\,\,\,\,\,\,\,\,\,\,\,\,
\,\,\,\,\,\,\,\,\,\,\,\,\,\,\,\,\,\,\,\,\,\,\,\,\,\,\,\,\,\,\,\,\,\,\,\,\,\,\,\,\,\,\,\,\,\,+h.a. \nonumber\\
&=&\frac{\hbar\Omega}{2} e^{i\delta t}\hat{J}_{x}-
\frac{\hbar\eta\Omega}{2}
e^{i(\delta+\omega_{0})t}\hat{a}^{\dagger}\hat{J}_{y}\nonumber\\
&&\,\,\,\,\,\,\,\,\,\,\,\,\,\,\,\,\,\,\,\,\,\,\,\,\,\,\,
-\frac{\hbar\eta\Omega}{2}
e^{i(\Delta-\omega_{0})t}\hat{J}_{y}\hat{a}+ h.a.
\end{eqnarray}
In this equation $\hat{J}^{(+)} =\hat{J}_{x}-i\hat{J}_{y}$ 
is the collective raising operator for the ions (i.e. the sum
$|2\rangle\langle 1|_{A}+|2\rangle\langle 1|_{B}$  of operators for the two ions A and B)
and, as before,  $\Delta$ is the detuning of the laser beam from 
the resonance frequency of the two-level system.  
Applying our formula eq.(\ref{zeta}) for the effective Hamiltonian we find
\begin{eqnarray}
\hat{H}_{eff}&=& 
\frac{\hbar\Omega^{2}\eta^{2}}{4 (\delta+\omega_{x})}
\left[\hat{J}_{y}\hat{a}, \hat{a}^{\dagger}\hat{J}_{y}\right]+
\frac{\hbar\Omega^{2}\eta^{2}}{4 (\delta-\omega_{x})}
\left[\hat{a}^{\dagger}\hat{J}_{y},\hat{J}_{y}\hat{a}\right] 
\nonumber\\
&=&\frac{\hbar\Omega^{2}\eta^{2}}{4(\delta-\omega_{x})}
\left(\frac{2\omega_{x}}{\delta+\omega_{x}}\right)
\hat{J}^{2}_{y}.
\label{constance}
\end{eqnarray}
This interaction is equivalent to a conditional quantum logic gate
preformed between the two ions, and can be used to create 
multiparticle entangled states. 

\section{Conclusion}
\setcounter{equation}{0}
This paper has presented a simple derivation of a useful compact formula, eq.(\ref{zeta}), for the effective Hamiltonian for course-grained or time-averaged dynamics.  We have found this formula to have been of considerable utility in a number of situations, and have presented some examples here.  We would welcome any comments from readers on other applications of this technique.  A more detailed exposition, with a more rigorous analysis of the validity of the formula is in preparation \cite{dfvj}.

\section*{Acknowledgements}
We would like to express our thanks to Sara Schneider, Timo K\"orber, Christian Roos, Enrique Solano and Pavel Lougovski for useful comments and correspondence. J. Jerke would like to thank the organizers of the 2000 Los Alamos Summer School, and thank the National Science Foundation and the Department of Energy for financial support.  This work was supported by the Los Alamos LDRD program and the Natural Science and Engineering Research Council of Canada.

\end{document}